# Aspects of strong electron-phonon coupling in superconductivity of compressed metal hydrides MH$_6$ (M=Mg, Ca, Sc, Y) with *Im-3m* structure

Pavol Baňacký[*] and Jozef Noga

Comenius University, Faculty of Natural Sciences, Chemical Physics group, Department of Inorganic Chemistry, Mlynska dolina CH2, 84215 Bratislava, Slovakia

**Abstract**   Recently, YH$_6$ was synthesized as a first compound from theoretically predicted stable compressed MH$_6$ hydrides with bcc *Im-3m* crystal structures. Superconductivity of pressurized YH$_6$ was confirmed with critical temperature (T$_c$) that is considerably lower than the predicted value by Migdal–Eliashberg (ME) theory. Here, we present theoretical reinvestigation of the superconductivity for selected MH$_6$ hydrides. Our results confirm that YH$_6$ and ScH$_6$ with *Im-3m* structure at corresponding GPa pressures are superconductors but with an anti-adiabatic character of superconducting ground state and a multiple-gap structure in one-particle spectrum. Transition into superconducting state is driven by strong electron-phonon coupling with phonons of H atom vibrations. Based on anti-adiabatic theory, calculated critical temperature T$_c$ in YH$_6$ is $\approx$ 231 K, i.e. just by $\approx$7 K higher than the experimental value. For ScH$_6$ the calculated critical temperature is T$_c$ $\approx$ 196 K. This value is by 27 K higher than a former theoretical prediction. Unexpected results concern CaH$_6$ and MgH$_6$ in *Im-3m* structure at corresponding GPa pressures. Calculated band structures (BS) indicate that in CaH$_6$ and MgH$_6$ the couplings to H stretching vibrations do not induce transitions into superconducting anti-adiabatic state and these hydrides remain stable in adiabatic metal-like state, which contradicts to former predictions of ME theory. These discrepancies are discussed in association with BS structure and a possible role of d-orbitals on the involved metals, while we stress that the anti-adiabatic theory uses BS topology and its stability as a key input.

# 1. Introduction

Increasing availability of computer power and development of robust effective optimization methods for *a priori* crystal structure prediction[1-3] in combination with DFT method [4,5] incorporated into some effective computer code for electronic structure calculation of periodic solids[6-8], opened a new area for possibility of uncovering unknown crystal structures of single elements, as well as of complex binary and ternary compounds just by specification of external pressure, temperature and stoichiometry of constituting elements in unit cell. Once the optimized structure, i.e. equilibrium geometry corresponding to minimum free energy for the involved external parameters is reached, electronic and phonon band structures are calculated on DFT level along with Eliashberg spectral function (ESF) - $\alpha^2F(\omega)$. The latter characterizes electron-phonon (EP) interactions and is directly related to EP coupling constant (EPC) - $\lambda$. Superconductivity research is a field where this —today rather routine, though sophisticated calculation "machinery"— is extensively exploited. As soon as EPC $\lambda$ is obtained, critical temperature $T_c$ can be straightforwardly calculated by simple semiempirical McMillan (MM) equation [9] or by its modification, namely the Allen-Dynes (AD) equation [10, 11] as derived for limiting range of $\lambda$ ($0 < \lambda < 1.5$) and pseudopotential $\mu^*$. Direct numerical solution of Eliashberg (EG) equations for superconducting gap ($\Delta$) within the Migdal-Eliashberg (ME) theory [12,13] is often applied for systems with $\lambda > 1$. The ME theory is a many-body extension of original weak coupling BCS theory[14]. However, it is also tacitly considered to be a strong coupling extension of the BCS. As pointed by, e.g. Marsiglio and Carbotte[15], ME theory is an extension of BCS theory for retardation effects due to much delayed phonon response at EP interactions. In this sense, it is still a weak coupling theory since characteristic electronic energy (e.g.. Fermi energy, $E_F$) is still dominant energy scale in a system.

In recent decade, hunt for room temperature superconductors[16] dominates in applications of the aforementioned calculation procedure. Very attractive and promising are polyhydrides that at high enough pressure basically simulate metallization of hydrogen. According to original idea of Ashcroft[17], the latter should be an ideal conventional BCS superconductor based on EP interactions with Cooper pair formation in k-space. Pressurized $H_2S$, i.e. $\{(H_2S)_2H_2\}$ with *Im-3m* structure was the first experimentally prepared high-pressure H-rich superconductor of this kind[18]. Its $T_c \approx 203$ K at 200 GPa was the highest $T_c$ recorded till then. Using the aforementioned calculation approach, the superconductivity of this species was predicted in 2014[19], i. e. a year prior to its synthesis. For $\lambda=2.19$ and $\mu^*$(0.1-0.13), based on AD equation, $T_c$ calculated[19] for *Im-3m* phase at 200 GPa was calculated in the range of 191 K–204 K, remarkably close to the experimental value[18].

Critical temperature $T_c$ over 200 K for compressed polyhydride was predicted already in 2012 by Wang et al.[20] for stable bcc *Im-3m* structure of $CaH_6$ with Ca atom encapsulated in a "cage" shaped by hexagons of 24 H atoms. Superconducting $T_c$ was derived within ME theory by numerical solution of isotropic EG equations in the range of 220 K–235 K at 150 GPa with $\lambda=2.69$. This prediction was confirmed by independent theoretical study[21] in 2017. For Mg, stable superconducting $MgH_6$ of bcc *Im-3m* high-pressure structure was predicted with even higher critical temperature[22]. Application of AD equation with $\lambda=3.29$ resulted in $T_c =263$ K at 300 GPa. Decrease of $T_c$ was predicted with increasing pressure in the range of 300 - 400 GPa. Numerical solution of EG equations in that range resulted in a similar character of pressure dependence but with $T_c$ well above the room temperature[23] (420 K–386 K).

Great expectations have been associated with predicted stable compressed polyhydrides of rare earth elements, in particular with "cage-like" fcc $Fm3m$ structure of $LaH_{10}$ and $YH_{10}$ and with $YH_6$ in bcc $Im-3m$ structure[24-26]. For $LaH_{10}$, based on solution of EG equations, calculated $T_c$ was 274 K–286 K at 210 GPa and for $YH_{10}$ predicted $T_c$ has been above room temperature reaching 305 K–326 K at 250 GPa, or 274 K -286 K at 210 GPa[26]. Based on MM equation, calculated $T_c$ for $YH_{10}$ is considerably smaller, 219 K-238 K at 210 GPa. In 2019, experimental realization of $LaH_{10}$ (fcc $Fm3m$ structure) was reported[27] with record-high detected critical temperature, $T_c$ = 250 K at 170 GPa, somewhat lower than predicted. Nevertheless, $T_c$ of 250 K was the highest experimentally reached $T_c$ until pressurized carbonaceous sulfur hydride {($H_2S$ +$H_2$)+$CH_4$} was synthesized[28] with the room-temperature $T_c \approx$ 288K at 267 GPa. No successful synthesis of $YH_{10}$ has been reported, as yet.

On the other hand, successful synthesis of stable $YH_6$ with bcc $Im-3m$ structure was announced recently[29,30]. Experimental value[29,30] of $T_c \approx$ 224 K at 166 GPa is substantially lower than the prediction[31] published only few months before its synthesis. By fully anisotropic ME theory, $T_c$ =290 K at 300 GPa was calculated[31] for $\lambda$ =1.73 and $\mu^*$=0.11. Former results for $YH_6$[32-35] are in the range of 165 K-264 K at 120 GPa with stronger EPC, $\lambda \approx$ 2.9-3.19.

Theoretical predictions of superconductivity were published also for Sc polyhydrides. Stable bcc $Im-3m$ structure and superconductivity of $ScH_6$ are predicted[36] at 285 GPa with $T_c$ = 130 K-147 K by AD equation for $\lambda$ =1.33. Prediction[24] of $T_c$ = 90 K -100K at 300 GPa with $\lambda$ =1.2 was based on numerical solution of EG equations. In 2018, Ashcroft and collaborators published[37] theoretical results for $ScH_6$ at 350 GPa. The AD equation predicted superconductivity at $T_c$ =135 K for $\lambda$ =1.2, whereas numerical solution of EG equations resulted in $T_c$ of 169 K. Unlike $YH_6$, high-pressure synthesis of $ScH_6$ with bcc $Im-3m$ structure has not been announced, so far.

It is difficult to imagine a practical application of the mentioned polyhydrides which are merely stable at such extreme pressures. On the other hand, these superconductors with remarkably high $T_c$ ( > 200 K) raise a serious need to ask whether current theoretical understanding of "strong coupling" superconductivity is adequate. In this respect, group of high-pressure metal hydrides $MH_6$ (M= Mg, Ca, Sc, Y) with the same bcc $Im-3m$ structure and superconductivity in exact stoichiometry without doping, is an ideal testing set for theoretical treatments. The ME theory predicts superconductivity with very high $T_c$ for all members of this group. Common feature behind high $T_c$ of these hydrides are EP interactions with EPC constant $\lambda$ > 1 (1.2 – 3.19). It is obvious that these values of $\lambda$ do not correspond to weak EP interactions, which is crucial assumption in formulation of BCS[14] theory and its extension, e.g. ME theory.

In this paper we present results of theoretical reinvestigation of superconductivity in specified $MH_6$ hydrides. Study is based on anti-adiabatic theory of superconducting state transition and aspects relevant for $MH_6$ are briefly introduced in the Supplementary Material. After synthesis of pressurized $H_2S$[18] with $T_c$ =204 K, the anti-adiabatic theory was successfully applied[38] to interpret the superconductivity and a high $T_c$ in this sulfide. Since the mentioned $MH_6$ hydrides —similarly as the compressed $H_2S$— are also characterized by a strong EPC with $\lambda$ > 1, there is a justified expectation that superconductivity in $MH_6$ hydrides is also associated with adiabatic → anti-adiabatic state transition. Obtained results confirm this expectation for $YH_6$ and $ScH_6$. Calculated $T_c$ is 231.5 K for $YH_6$ and 195.7 K for $ScH_6$. Calculated $T_c$ for $YH_6$ is 7.5 K higher than experimental $T_c$. Prediction of ME theory is higher by more than 60K. For $ScH_6$, $T_c$ calculated by anti-adiabatic theory is by 26.7 K higher than the former ME prediction. Surprising results are obtained, however, for bcc $Im-3m$ structure of the studied alkaline earth metal hydrides. Electronic band structure calculations indicate that EP interactions in $CaH_6$ and $MgH_6$, in particular coupling to H stretching vibrations, do not induce transition from adiabatic state into superconducting anti-adiabatic ground state. These hydrides remain stable in adiabatic metal-like state characteristic for frozen equilibrium structure within the framework of Born-Oppenheimer approximation. It is in sharp contradiction with ME theory, which for both compounds predicts superconductivity with extremely high $T_c$. Possible reason of this contradiction with the predictions of ME theory is discussed in Sec. 4.

## 2. Remarks to strong EP interactions; small Fermi energy physics

There are no clear specifications that $\lambda > 1$ is compatible with ME theory[12,13]. Range of validity of generic BCS theory and BCS-like theories for EP interactions were specified by Migdal[12] and Eliashberg[13]. The matter is very complex to be analyzed and discussed here. What is important for our study can be formulated roughly as follows; Migdal's theorem (neglecting vertex corrections) requires validity of the condition $\omega\lambda/E_F << 1$ and Eliashberg restricts validity of this inequality only for $\lambda \leq 1$. From these inequalities directly follows a natural requirement that BCS and BCS-like theories (including ME theory) are valid only for adiabatic systems which obey Born–Oppenheimer approximation (BOA). Expressed explicitly, $\omega/E_F << 1$ must hold for characteristic energy scales ($\omega$, $E_F$) of interacting phonon and electronic subsystems. This is perfectly fulfilled in study superconductivity of elementary metals and theirs alloys, for which BCS as well ME theory were basically developed. The adiabatic BOA is crucial approximation of theoretical molecular and solid state physics. Adiabatic ratio increased on the level $\omega/E_F < 1$ brings nontrivial problems into theory. It indicates importance of non-adiabatic contributions in calculation of EP interactions within the BOA, an effect which is beyond standard ME approach. Attempt to solve this problem was formulation of the non-adiabatic theory of superconductivity by Pietronero and coworkers[39,40]. Non-adiabatic theory has solved this nontrivial problem by generalization of EG equations through accounts for vertex corrections and cross phonon scattering which is beyond standard ME approximation. The theory, non-perturbative in $\lambda$ and perturbative in $\omega\lambda/E_F$, has been applied for interpretation of different aspects of high-$T_c$ superconductivity[41]. It accounts for non-adiabatic effects in a quasi-adiabatic state $\omega/E_F \leq 1$ and is able to simulate various properties of high-$T_c$ superconductors, including high-values of $T_c$, for moderate EPC, $\lambda \approx 1$. In addition, it has also been shown that increased electron correlation is an important factor which makes corrections to vertex function positive, which is crucial for increasing $T_c$.

Nonetheless, sophisticated treatment of high-$T_c$ superconductivity within the non-adiabatic theory faces serious problem related to possibility of non-adiabatic polaron collapse of the band and bipolaron formation. However, polaron collapse is associated with the translation symmetry breaking. This fact raises serious questions about formulation of ME theory that is based on Green functions method which assume translational invariance. According to bipolaron theory of Alexandrov[42-46], polaron collapse occurs already at $\lambda \approx 0.5$ for uncorrelated polarons and even at a smaller value for bare EPC in strongly correlated systems. For $\omega/E_F \leq 1$, or $\lambda \geq 1$ and for $\omega/E_F \geq 1$ and any small value $\lambda<<1$, the non-adiabatic polaron theory has been shown to be basically exact[46]. Bipolarons can be simultaneously small and light in a suitable range of Coulomb repulsion and EP interaction[47]. These results have important physical consequences. There are serious arguments that an effect of polaron collapse cannot be covered through calculation of vertex corrections due to translation symmetry breaking and, mainly, polaron collapse changes a possible mechanism of pair formation. Instead of BCS scenario with Cooper pair formation in momentum space, the Bose-Einstein condensation (BEC) with mobile bipolarons (charged bosons) in real space is realized.

Unfortunately, physics behind a microscopic mechanism of superconductivity is even more complicated as sketched above. In 2004/2005 two independent studies[48,49] of EP interactions in 40 K superconductor $MgB_2$, with intermediate EPC $\lambda=0.87$, were published. It was shown that within the root-mean square displacement ($q_{rms}$) of B atoms in vibration ground state (zero-point energy), degeneracy of σ bands in Γ is lifted and maximum of one of the bands starts to fluctuate. This band approaches the Fermi level (FL) within an energy range that is smaller than the vibration energy (ω), and, with increasing vibration displacement the band maximum merges below the FL. Motion of this fluctuating band across the FL follows the B-B vibration motion. This trivial effect induces crucial physical consequences. First of all, Fermi energy ($E_F$) is no more constant. It contradicts to basic assumption in the formulation of BCS theory. With increasing vibration displacement, $E_F$ decreases toward values of phonon energies (ω), then continues below these values and, when the top of the fluctuating band approaches FL, $E_F$ approaches 0 eV. Decrease of $E_F$ means that originally adiabatic (or quasi-adiabatic) state, $\omega/E_F \ll 1$ (or $\omega/E_F < 1$), becomes non-adiabatic ($\omega/E_F \approx 1$) and finally undergoes transition into anti-adiabatic state, $\omega/E_F > 1$. Transition is induced by EP coupling to $E_{2g}$ optical stretching mode in $MgB_2$. This situation, i.e. small Fermi energy physics, is beyond theory of metals and also beyond BCS and ME theories and, what is crucial, it is beyond the adiabatic BOA. Theoretical study of $MgB_2$ superconductivity in anti-adiabatic state has been presented in Ref [48]. Study is based on non-adiabatic theory of electron–vibration interactions beyond the BOA. Basic information on non-adiabatic theory of electron–vibration interactions and methods used are available online in Supplementary material.

Present study of EPC and superconductivity in compressed $MH_6$ has identified the importance of small Fermi energy physics also in these hydrides.

## 3. Results

Studied hydrides $MH_6$ (M=Y, Sc, Mg, Ca) are stable in bcc *Im-3m* crystal structure at individual – metal-specific GPa pressure and lattice parameter *a*. Crystal structure is characteristic by a "cage" shaped by hexagons of 24 H atoms which encapsulate the metal atom M. Common feature of these hydrides is the dominance of $E_g$ (2-fold degeneracy) and $T_{2g}$ (3-fold degeneracy) optical phonon modes of H vibrations[32]. High-frequency optical phonon modes of H vibrations contribute roughly by 80% to total EPC value λ. In Fig. 1 we show the bcc *Im-3m* crystal structure with one H-hexagon marked by yellow corona and with arrows indicating displacements in stretching vibration of H-atoms in $E_g$ mode. To simulate $E_g$ stretching vibration in this "cage" structure by a model of frozen stretching mode, it is advantageous to use *cP* unit cell (u.c.) with 2 formula units (f.u.). It is justified by the fact that topology/position of $MH_6$ bands which are nearest to FL in zone-center Γ point in *cI* lattice with 1 f.u./u.c. is the same as for *cP* lattice with 2 f.u./u.c. (cf., Γ- point bands degeneracy in Figs. 2 and 3 for equilibrium geometry). Our interest is focused on lifting of the band degeneracy in Γ point with H atoms displacement. In this case, the H atoms displacements $\delta$ within the unit cell with 2 f.u. obey following fractional coordinates scheme: H1 ½+$\delta$,¼,0; H2 ¼-$\delta$, ½,0; H3 ½+$\delta$, ¾,0; H4 ¾-$\delta$, ½,0;

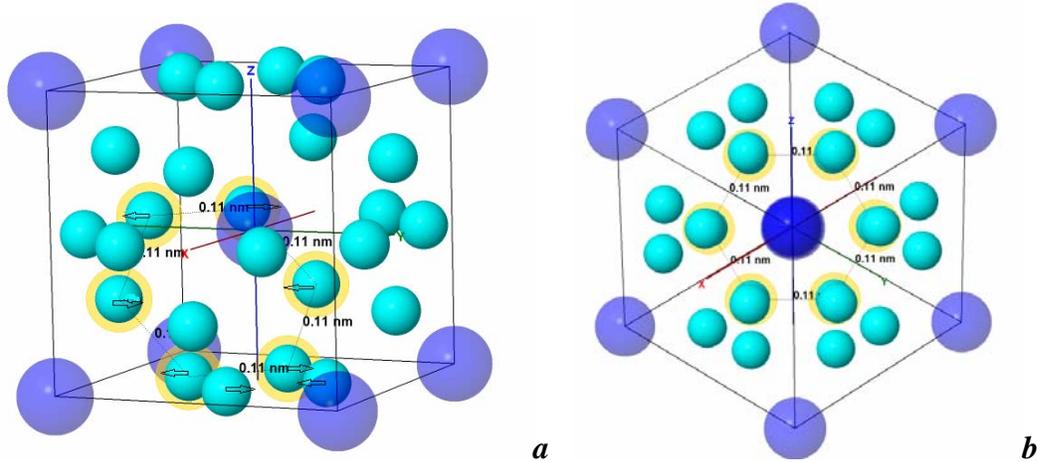

**Fig. 1** Crystal structure bcc *Im-3m* of MH$_6$ hydrides for metal-specific GPa pressures. In particular, as an example: *a*/ shown is the MgH$_6$ structure at 300 GPa with H-H distance 1.1 Å as indicated in one of the hexagons with H atoms decorated by a yellow corona. The H atoms are small cyan balls and larger blue balls are Mg atoms. Arrows in decorated H atoms of hexagon indicate displacements in E$_g$ mode stretching vibration; *b*/ displayed is the view of the structure *a*/ in [111] direction - some of H atoms are eclipsed.

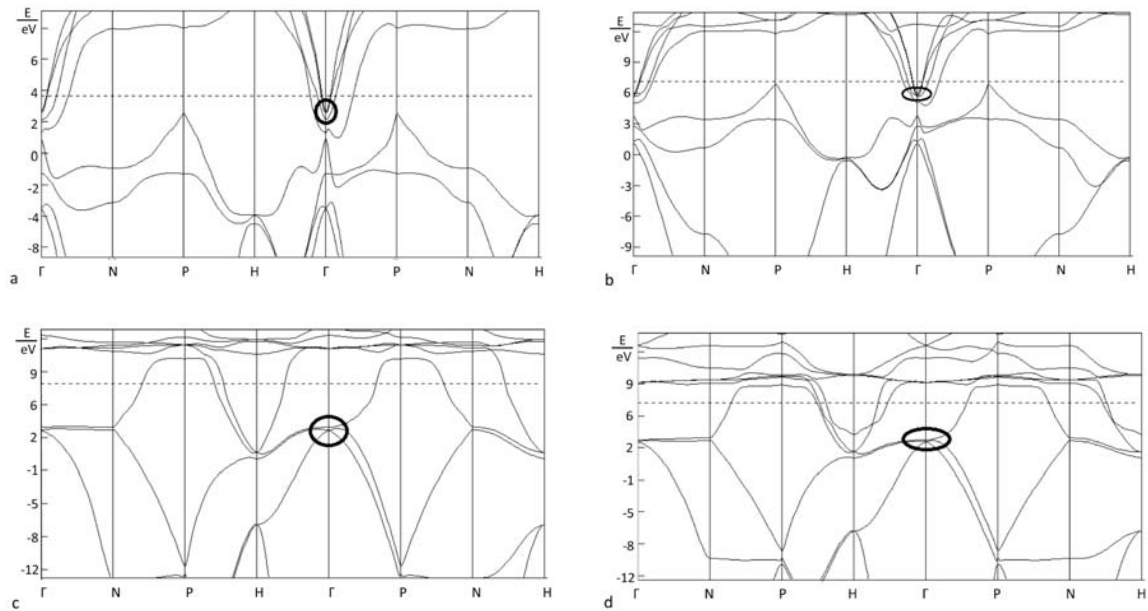

**Fig. 2** Electronic BS in *cI* lattice calculated for the investigated set of compressed hydrides which are stable in Im-3m structure at metal-specific GPa pressures – a/ YH$_6$ (a=3.578 Å), b/ ScH$_6$ (a=3.266 Å), c/ MgH$_6$ (a=3.1116 Å), d/ CaH$_6$ (a=3.501 Å). Cyclic cluster HF-SCF method was used for the calculations (see Supplementary material S2) for *cI* Im-3m lattice with 1 f.u./ u.c. For YH$_6$ and ScH$_6$, u.c. was treated as an open-shell system by UHF procedudure. MgH$_6$ and CaH$_6$ with closed-shell u.c. were calculated by RHF procedure. For the determination of superconducting properties, most important are the bands degenerated in zone-center Γ point.

H5 ¼+δ,0, ½; H6 ½-δ,0,¼; H7 ¾+δ,0,½; H8 ½-δ,0,¾; H9 0, ¼+δ, ½; H10 0, ½-δ, ¾; H11 0, ¾+δ, ½; H12 0, ½-δ, ¼; M13 0,0,0; M14 ½,½,½. Post-SCF calculations of non-adiabatic corrections in a system beyond the BOA, which are needed for determination of superconductivity parameters within the anti-adiabatic theory, are based on data extracted from the BS structure calculations.

**Yttrium hydride YH$_6$**

We have studied electronic BS for experimental[29] and also for predicted[31] lattice parameter *a*. Experimental[29] value is *a*=3.578 Å at 166 GPa and *a*=3.369 Å was predicted[31] for equilibrium *Im-3m* structure at 300 GPa. Topology of BS dispersion for Γ-X path ([000] – [0½0]) close to FL is crucial in the investigation of stability/instability of the electronic structure due to a coupling with optical stretching mode. Simulation of stretching vibration by frozen mode method has been performed in cP lattice with 2 f.u./u.c.

BSs calculated for experimental (*a*=3.578Å) equilibrium geometry and for distorted geometries with different displacements *δ* of H atoms in stretching vibration mode E$_g$ of YH$_6$ are shown in Figs. 3a-c. For predicted - optimized[31] value of *a*=3.369Å, the BS was calculated with basically the same character of dispersion and topology of bands. It can be seen that EPC of E$_g$ mode induces splitting (Figs. 3b,c) of upper-most valence band minimum in Γ point that is triply degenerated at fixed equilibrium geometry–Fig 3a. It should be noted that in equilibrium geometry this system is quasi-adiabatic. Fermi energy is E$_F$≈0.54 eV and, for phonon mode stretching frequency ω$_{HH}$=1650cm$^{-1}$ (49.47 THz; 204.8 meV), the adiabatic ratio is ω$_{HH}$/E$_F$≈0.38. Displacements of H atoms out of equilibrium simulate stretching vibrations. Figs. 3b,c show that vibrations not only induce a splitting of the band-degeneracy in Γ point but the bands become fluctuating, as well. For *δ*=0.024 (in fractional units - fr.u.), minimum of the band is shifted close to FL – Fig. 3b and still increased vibration displacement to *δ*=0.025 shifts the band-minimum above FL – Fig.3c. Band shift is periodic and follows vibration displacements. It is obvious that this effect indicates electronic structure instability related to breakdown of the BOA. When the energy difference of the minimum of the fluctuating band w.r.t. FL in Γ point is less than ω$_{HH}$ (abs ε$_{min}$ < ω$_{HH}$, i. e. either from valence or conducting side of FL), the system is in anti-adiabatic regime. It means that effective velocity of electrons in this region is slow-downed and dynamics of nuclei (ions) becomes important, i.e. ω / E$_F$ > 1. Extreme situation should occur when minimum of the band (in general it is an analytical critical point of band - ACP) coincides with FL and E$_F$ decreases to 0 eV.

For experimental parameter *a*=3.578 Å at 166 GPa, we have calculated the FL crossing by ACP of the fluctuating band for H-vibration displacement *δ$_{cr}$*= 0.0247fr.u.. In stretching vibration, H atom oscillates ± *δ$_{cr}$* around equilibrium position, i.e. the total vibration displacement is 2δcr= 0.1767 Å. In order to justify validity of the BOA for this system, the ACP of fluctuating band has to stay below the FL for H atom total vibration displacement that is larger than that corresponding to the root-mean square displacement of the vibration ground state - $q_{rms}$. For H-stretching vibration $\omega_{HH}$=1650cm$^{-1}$ (49.47 THz; 204.8 meV), $q_{rms}$ = 0.1791 Å, as calculated at the level of LHO approximation (Eq. S2.2 in Supplementary material). This result indicates that YH$_6$ at 166 GPa is unstable in EPC to H-stretching vibrations. The BOA is broken and the system has to be studied in anti-adiabatic state by non-adiabatic electron-vibration theory beyond the BOA. Qualitatively equivalent results were obtained for optimized *a*=3.369 Å at 300 GPa. In this case, the calculated critical $\delta_{cr}$= 0.0247fr.u. and the total vibration displacement was 2$\delta_{cr}$= 0.1684 Å. For stretching vibrations in the range of $\omega_{HH}$=1613 cm$^{-1}$– 1775 cm$^{-1}$(48.36 - 53.2 THz; 200 - 220 meV), the LHO approximation value for $q_{rms}$ varies from 0.1811 Å to 0.172 Å. Again, the system is unstable in EPC to H-stretching vibration and BOA is broken.

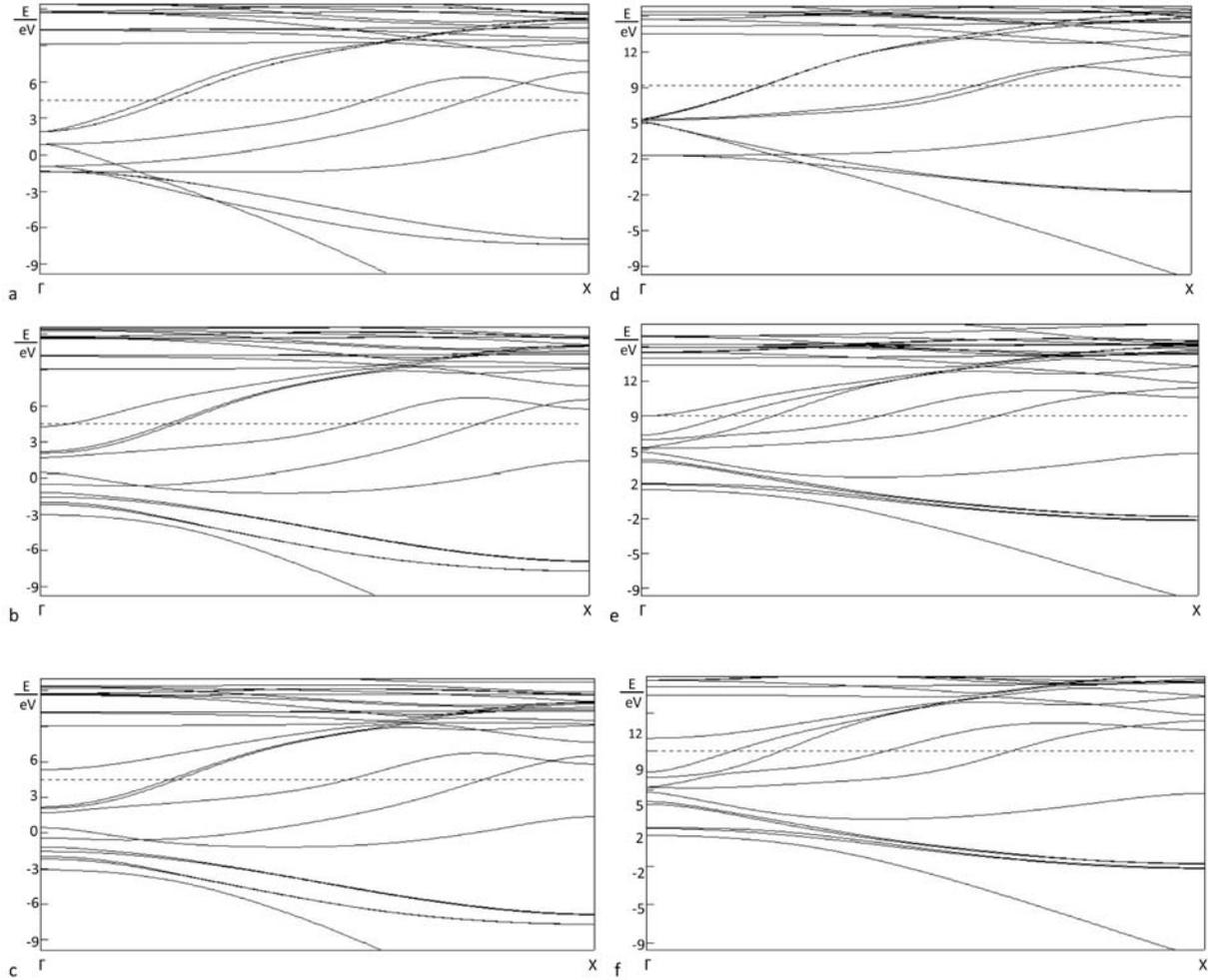

**Fig. 3** Topologies of BS dispersion for **YH$_6$** (**a/** equilibrium geometry; H-displacements in E$_g$ vibration mode **b/** δ=0.024; **c/** δ=0.025) and **ScH$_6$** (**d/** equilibrium geometry; H-displaced structures **e/** δ=0.018; **f/** δ=0.019). Displacements are in fractional units (fr.u.).

System can be stabilized in anti-adiabatic state if non-adiabatic correction $\Delta E^0_{(na)}$ (see S2.3) to its adiabatic electronic ground state energy is negative, $\Delta E^0_{(na)} < 0$, and in absolute value this correction is greater than an energy loss (decrease) due to H-displacements, $\Delta E_{\delta cr} > 0$. In short, for critical displacement $\delta_{cr}$, $abs(\Delta E^0_{(na)}) > \Delta E_{\delta cr}$ has to hold. Here, $\Delta E_{\delta cr}$ is the contribution of nuclear potential energy at displacement $\delta_{cr}$ to a total electronic energy on adiabatic level. Inspection of Eq. (S2.3) indicates that crucial quantities for this correction are PDOS related to bands close to FL and matrix elements $u^r_{k-k'}$ of electron-vibration interactions. From Eq. (S2.3) also follows that the contribution $\Delta E^0_{(na)} < 0$ can merely originate from EP interactions of electronic states in fluctuating band when ACP approaches FL with remaining 4 electronic bands that intersect FL – Figs. 3b,c. Calculated PDOS, $n^0(\varepsilon^0_k) = |\partial \varepsilon^0_k / \partial k|^{-1}$, of these 4 bands at FL are quite small. In average, PDOS for each band is ≈ 0.03 states/eV. On the other hand, shift of the ACP of fluctuating band toward FL increases the PDOS of this band at FL remarkably – see van Hove–like PDOS appearance in Fig. 4.b. calculated for dispersion of $YH_6$ in Fig.4.a. For energy position of the ACP equal to $\omega_{HH} / 2 \approx 100$ meV with respect to FL, mean value of PDOS is ≈ 0.15 states/eV.

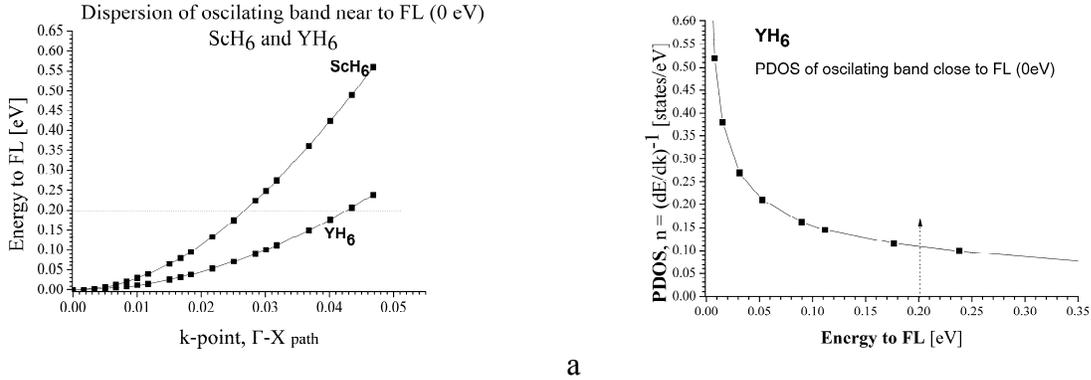

**Fig.4 a/** Comparison of band dispersion near Γ point of fluctuating bands in $YH_6$ and $ScH_6$ at displacement when band minimum approaches FL. As an ilustration, energy position of fluctuating band minimum w.r.t. FL equals to $E_g$ stretching energy $\omega_{HH}$ is indicated by the horizontal dotted line. Energy of FL is rescaled to 0 eV. In **b/** is PDOS of fluctuating band in $YH_6$ close to FL. Energy to FL indicated by vertical arrow corresponds to $\omega_{HH}$ energy value. Increased PDOS of this band at FL is induced by EPC due to H-vibrations and corresponds to $YH_6$ dispersion displayed in Fig.4a. In equilibrium geometry PDOS of this band at FL is 0 – band minimum is below FL more than $\omega_{HH}$ (cf. Fig 3a).

The BS calculations are based on semiempirical INDO Hamiltonian (see Supplementary material S2.). Accordingly, in post-SCF calculation of non-adiabatic energy corrections (S2.3-S2.5), instead of individual matrix elements of electron vibration interaction $u^r_{k-k'}$ it is adequate to use a mean value of this interaction - $g_{\delta cr}$. The latter is calculated by Taylor expansion of a mean value of one-electron Hamiltonian ($<h_1(r,Q)>$) with respect to atom displacement in particular stretching mode – see (S2.12). In contrast to dimensionless λ, the mean-value $g_{\delta c}$ is given in energy units, eV.

For YH$_6$ in experimental geometry, the mean value of EPC calculated in this way was $g_{\delta cr} \approx 3.6$ eV. For predicted-optimized geometry at 300 GPa, mean-value of EPC was $g_{\delta cr} \approx 3.7$ eV. Based on the calculated physical parameters, non-adiabatic corrections to adiabatic energies and resulting superconducting properties are as follows;

*a/ Stabilization energy in anti-adiabatic ground state;* $\Delta E_{(aa)} = \Delta E^0_{(na)} + \Delta E_{\delta cr}$.

For experimental adiabatic equilibrium geometry the calculated values are: $\Delta E_{\delta cr} \approx +0.38$ eV, $\Delta E^0_{(na)} \approx -1.86$ eV and the system is stabilized in anti-adiabatic state with distorted geometry by $\Delta E_{(aa)} \approx -1.48$ eV. For optimized adiabatic geometry at 300GPa, $\Delta E_{\delta cr} \approx +0.41$ eV, $\Delta E^0_{(na)} \approx -2.00$ eV and the system is stabilized in anti-adiabatic state with distorted geometry by $\Delta E_{(aa)} \approx -1.59$ eV. Stabilized anti-adiabatic ground state is geometrically degenerated – it has a fluxional structure in positions of H atoms distorted by displacements $\delta_{cr}$. The involved displaced nuclei can revolve in concert on perimeters of circles with radiuses $\delta_{cr}$, while the total electronic energy remains unchanged because system remains in stabilized anti-adiabatic ground state. Circles of revolving are centered in adiabatic equilibrium positions. On a lattice scale, network of alternating short and long "H-H bonds" evolves. Motion in concert means that in nuclear revolving on a lattice scale, H-H displacements preserve directional character of stretching mode. This enables creation of mobile bipolarons (i.e. mobile inter-site polarized charge density distribution in real space) and their motion in an external electric potential is without energy dissipation.

*b/ Gap formation in real space; non-adiabatic correction $\Delta \varepsilon_{Pk}$ to adiabatic band energies $\varepsilon^0_{Pk}$.*

Inspection of Eqs. (S2.4, S2.5) indicates that EP interactions of electronic states of the fluctuating band (in ACP position close to FL) with electronic states of next 4 bands that intersect FL do induce shifts of electronic states of these bands in a close vicinity of FL. Electronic states above FL ($\varepsilon^0_{k'}$ for $k'>k_F$) are pushed upward from FL by non-adiabatic correction (S2.4) - $\Delta \varepsilon_{k'}$. Electronic states which are close but below FL, ($\varepsilon^0_k$ for $k<k_F$) are pushed downward from FL by non-adiabatic correction (S2.5) - $\Delta \varepsilon_k$. Shifts of electronic states at FL naturally imply changes in PDOS at FL as compared to the original adiabatic one - $n^0(\varepsilon^0_k) = |(\partial \varepsilon^0_k / \partial k)|^{-1}$. Corrected PDOS are calculated according to equation (S2.6). The gap is identified as the energy difference between created density peaks in corrected PDOS above the FL ($\Delta(0)/2$ - half-gap) and below the FL in anti-adiabatic state. Formation of density peaks is related to the spectral weight transfer that can be observed by ARPES or tunneling spectroscopy in cooling below T$_c$. Theoretical study of spectral weight transfer-simulation of ARPES and gap formation in different superconductors we presented in Ref[48] and Ref[6,7(in SI)].

In actual situation (see Figure 3a,b), the fluctuating band interacts with 4 bands that intersect FL on Γ-X path. Close to FL (within the range of energy± $\omega_{HH}$ w.r.t. FL), the calculated adiabatic PDOS for these bands are basically the same and quite small, $n^0_{FL} \approx$ 0.03states/eV. According to former study[32], H-vibration modes contribute by 80 % to total EPC strength. It should be reminded that total degeneracy of H-vibration modes is 5. Next, we assume that EPC strength is the same for interaction with all 4 bands. It is straightforward to calculate that with these assumptions, the mean value of EP interaction strength is $g \approx 3.6$ eV/band for YH$_6$ in experimental geometry and $g \approx 3.7$ eV/band in case of optimized geometry, $a$=3.369 Å at 300 GPa. Orbital energy corrections, $\Delta\varepsilon_k$, ($\Delta\varepsilon_k$), have been calculated for 7 different positions of ACP of fluctuating band w.r.t. FL over relevant energy range (i.e.[10meV–200 meV] – see Figure 3b).

For experimental geometry ($a$=3.578 Å at 166GPa), in the $k$ point where the particular band intersects the FL, the mean value of opened gap is $\Delta(0) \approx$ 79.8 meV. In one-particle spectrum, such gap is opened in all 4 bands that intersect FL on adiabatic level (Figs. 3b,c). With optimized $a$=3.369 Å at 300 GPa, the calculated value of opened gap is $\Delta(0) \approx$ 87.4 meV.

At critical temperature, the gap(s) in one-particle spectrum is (are) closed (S2.8), $\Delta(T_c)$=0, and adiabatic continuum of band states is established at FL. In these circumstances $\left|\Delta E^0_{(na)}(T_c)\right| \leq \Delta E_{d,cr}$ holds and the system undergoes a transition from the anti-adiabatic into adiabatic state. The latter is stable for equilibrium high-symmetry structure above T$_c$. In adiabatic state, correction is small and positive $\Delta E^0_{(na)}$>0. It corresponds to an energy contribution of a standard adiabatic polaron that contributes to the total energy of the system by a small value, only (see Eq. S2.10).

*c/ Critical temperature $T_c$ of compressed* YH$_6$

Final T$_c$ was calculated according to equation (S2.9) - $T_c = \Delta(0)/4k_B$, based on the mean value of the calculated $\Delta(0)$. For the system in experimental geometry ($a$=3.578A at 166GPa), from anti-adiabatic theory follows T$_c \approx$ 231.5 K. In optimized geometry ($a$=3.369Å at 300 GPa), critical temperature is T$_c \approx$ 237.5 K.

**Scandium hydride ScH$_6$**

For BS calculation and study of superconducting properties we have used predicted[37] lattice parameter of stable *Im-3m* phase of ScH$_6$, $a$=3.266 Å. Calculated BS is presented in Figs. 3d-f. Alike as for YH$_6$, in fixed equilibrium geometry, the upper-most valence band minimum in Γ point is triply degenerated. In this case, too, the system is adiabatic with E$_F \approx$0.79 eV and for phonon mode stretching frequency $\omega_{HH}$=1600cm$^{-1}$ (47.97 THz; 198.4 meV) adiabatic ratio is $\omega_{HH}$/E$_F \approx$ 0.25. Displacements of H atoms out of equilibria, which simulate stretching vibrations, induce lifting of bands minima degeneracy in Γ point (cf. Figs 3e,f). For a critical displacement $\delta_{cr}$= 0.019 fr.u., the minimum, i.e. ACP of fluctuating band, is shifted above the FL (Fig. 3f). Corresponding total vibration displacement of H atom is $2\delta_{cr}$ = 0.124 Å. For stretching vibration $\omega_{HH}$=1600 cm$^{-1}$ (47.97 THz; 198.4 meV), the calculated $q_{rms}$ in LHO approximation is $q_{rms}$=

0.182 Å. For $\omega_{HH}$=1650 cm$^{-1}$ (49.47 THz; 204.8meV ), $q_{rms}$ in LHO approximation is $q_{rms}$ = 0.179 Å. Again, the system is unstable in EPC w.r.t. H-stretching vibrations and BOA is broken.

Also for $ScH_6$, shift of the ACP of fluctuating band toward FL increases PDOS for the fluctuating band at FL. In Figure 4a, dispersions of fluctuating bands for $ScH_6$ and $YH_6$ are compared. It is evident that at relevant energy distance $\pm\omega_{HH}$ to FL, adiabatic PDOS of $ScH_6$ should be smaller than PDOS of $YH_6$. This is confirmed by Figure 5, where PDOS of $ScH_6$ is displayed as a function of energy distance of the band-minimum to FL.

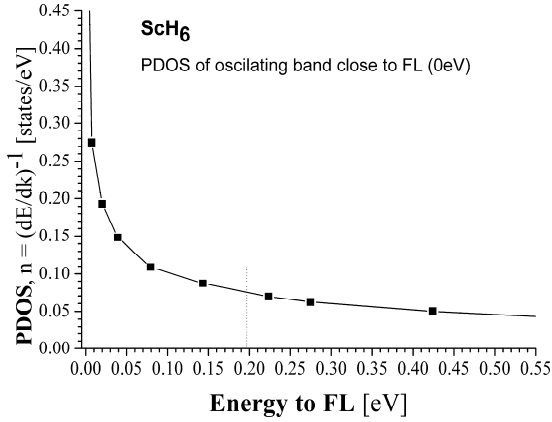

**Fig. 5.** PDOS of fluctuating band in $ScH_6$ close to FL. Energy corresponding to $\omega_{HH}$ energy value is indicated by the vertical dotted line. Increased PDOS of this band at FL is induced by EPC due to H-vibration motion. In equilibrium geometry, PDOS of this band at FL is 0 – band minimum is below FL (see Fig. 2b and Fig. 3d).

For energy range $\pm \omega_{HH} / 2 \approx 100$ meV of the ACP position w.r.t. FL, a mean value of PDOS is $\approx 0.1$ states/eV comparing to 0.15 states/eV for $YH_6$. Even smaller are PDOS of remaining 4 bands which intersect FL on the Γ-X path. Within $\pm \omega_{HH}$ to FL, PDOS of each band is $\approx 0.03$ states/eV. Mean value of electron–vibration interaction $g_{\delta cr}$ is calculated in the same way as for $YH_6$. In H-displacements, which induce FL crossing by ACP of fluctuating band, this interaction is stronger for $ScH_6$, namely, $g_{\delta cr} \approx 4.25$ eV. Based on physical parameters presented above, non-adiabatic corrections to adiabatic energies and resulting superconducting properties were calculated as follows:

*a/ Stabilization energy in anti-adiabatic ground state;* $\Delta E_{(aa)} = \Delta E^0_{(na)} + \Delta E_{\delta cr}$ .

The decrease (energy loss) of the total electronic energy due to critical H-displacements in stretching mode is $\Delta E_{\delta cr} \approx +0.21$ eV. Non-adiabatic correction is $\Delta E^0_{(na)} \approx -2.6$ eV and the system is stabilized in anti-adiabatic state by $\Delta E_{(aa)} \approx -2.39$ eV. Stabilized anti-adiabatic ground state is geometrically degenerated – it has a fluxional structure in position of H atoms on perimeters of circles with radii of $\delta_{cr}$ .

*b/ Gap formation in real space; non-adiabatic correction $\Delta\varepsilon_{Pk}$ to adiabatic band energies $\varepsilon_{Pk}^0$*

In k point where the particular band intersects FL, the mean value of opened gap is $\Delta(0) \approx 72$ meV for $\omega_{HH}=1600cm^{-1}$. For $\omega_{HH}=1650cm^{-1}$ (49.47 THz, 204.8meV), the calculated opened gap is $\Delta(0) \approx 73.4$ meV. In one-electron spectrum, the gap is open on the path Γ-X at k-points where the particular bands intersect FL on adiabatic level in equilibrium geometry (Figure 3f). For critical temperature $T_c$, gap(s) in one-particle spectrum is (are) closed (S2.8), $\Delta(T_c)=0$, and adiabatic continuum of band states is established at FL. In these circumstances $\left|\Delta E_{(na)}^0(T_c)\right| \leq \Delta E_{d,cr}$ holds, and the system undergoes a transition from the anti-adiabatic into adiabatic state which is stable for equilibrium high-symmetry structure above $T_c$. In adiabatic state, correction is small and positive, $\Delta E^0_{(na)} > 0$. This is basically a small energy contribution of a standard adiabatic polaron(s) to the total electronic energy of system (see Eq. (S2.10)).

*c/ Critical temperature $T_c$ of compressed $ScH_6$*

Final $T_c$ was calculated according to equation (S2.9) - $T_c = \Delta(0)/4k_B$ based on the mean value of calculated $\Delta(0)$. Critical temperature based on anti-adiabatic theory is $T_c \approx 195.7$ K for $\omega_{HH}=1600cm^{-1}$ and $T_c \approx 199.5$ K for $\omega_{HH}=1650cm^{-1}$.

**Magnesium hydride $MgH_6$**

Predicted[22], optimized stabile *Im-3m* structure for $MgH_6$ is for pressure over 263 GPa. At 300 GPa, the H-H distance is 1.1 Å and corresponding lattice parameter is $a$=3.111 Å. This is the shortest H-H distance among studied $MH_6$ hydrides. This distance is nearly as short as distance predicted for metal hydrogen at 500 GPa in Cs-IV phase. As it could be expected for *Im-3m* structure of $MgH_6$, strong charge transfer and strong EPC has been calculated[22], $\lambda\approx3.29$. Mentioned parameters are in support of high Tc. Application of ME theory, either AD equation or numerical solution of EG equations, confirmed[22,23] these expectations as presented in the Introduction.

However, application of non-adiabatic theory of electron-vibration interaction in the present study does not confirm that $MgH_6$ –stable in *Im-3m* structure over 263 GPa– is a superconductor as predicted by ME theory. In Fig. 6 we present the calculated BS for distorted structures of both alkaline earth metal hydrides. BS in Fig 6a corresponds to $MgH_6$ in distorted geometry with H atoms displacement of $\delta$=0.04 fr.u., i. e. the total vibration displacement of an H atom was $2\delta$ = 0.249 Å. For $\omega_{HH}=1600$ cm$^{-1}$, in LHO approximation $q_{rms} \approx 0.182$ Å and for $\omega_{HH}=1650$ cm$^{-1}$, $q_{rms}$ = 0.179 Å. It is obvious from Fig. 6a that BS of $MgH_6$ remains stable even for such a large H-displacemet (c.f. Fig. 6a and Fig. 2c). In Γ point, the band-degenracy is lifted but fluctuating band is situated deeply below FL and the system does not undergo transition to superconducting anti-adiabatic ground state. $MgH_6$ remains with metallic character in adiabatic state with $E_F\approx2.5$eV. The mean value of electron–vibration interaction calculated for a displacement $\delta$ =0.0181 fr.u. is $g_\delta \approx 2.43$eV. This value is relatively strong, though smaller than $g_{\delta cr} \approx 4.26$ eV calculated for ScH$_6$ with almost equivalent $\delta_{cr}$ of 0.019 fr.u. Energy loss of total electronic energy due to H-displacements in stretching is $\Delta E_\delta \approx +0.2$ eV which is nearly identical to $\Delta E_{\delta cr}$ in ScH$_6$, +0.21 eV.

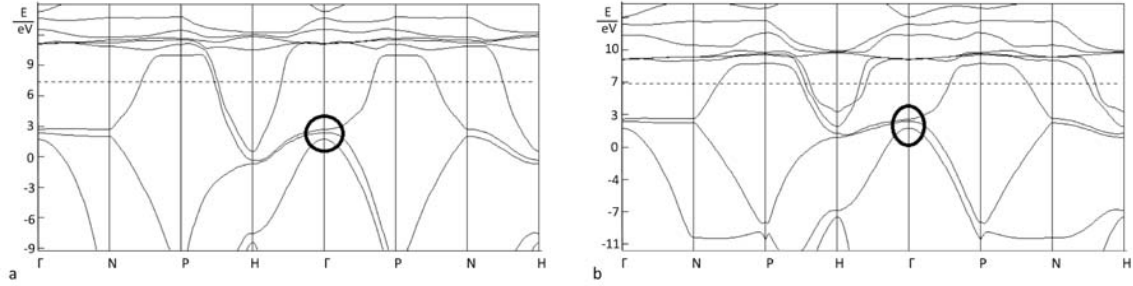

**Fig.6** BS calculated for distorted structures of **a**/ MgH$_6$ and **b**/ CaH$_6$ with H-displacement 0.04 fr.u. in cI Im-3m lattice. Compare to equilibrium structures in Figure 2c,d – see the text

**Calcium hydride CaH$_6$**

Like for MgH$_6$, application of non-adiabatic theory of electron-vibration interaction in our current study of CaH$_6$ did not confirm that this hydride –stable in *Im-3m* structure at 150 GPa– is superconductor as predicted[20,21] by ME theory. Our results indicate that the system is stable in coupling to H-stretching vibration within the BOA. In Fig. 6b, the BS for CaH$_6$ calculated in distorted geometry with H atoms displacement of $\delta = 0.04$ fr.u. (it corresponds to total vibration displacement of H atom $2\delta = 0.28$ Å) is presented. Root-mean square displacement in LHO approximation for $\omega_{HH}$=1600 cm$^{-1}$ is $q_{rms} \approx 0.182$ Å and for $\omega_{HH}$=1650 cm$^{-1}$, $q_{rms} = 0.179$ Å. It is obvious from Fig. 6b that BS of CaH$_6$ remains stable even for 0.28 Å H-displacemets (cf. Fig. 6b and Fig. 2d). In Γ point the bands degeneracy is lifted, but fluctuating band is deeply below FL and the system does not undergo a transition to superconductiong anti-adiabatic ground state. CaH$_6$ remains in metal-like adiabatic state with E$_F \approx$2eV, in spite of a relatively strong EP interactions of $g_\delta \approx$2.66eV.

## 4. Discussion and Conclusion

Application of non-adiabatic theory of electron-vibration interaction in study of high-Tc superconductivity of compressed MH$_6$ hydrides with *Im-3m* structure indicates that theoretical understanding of strong EPC in superconductivity phenomenon still remains an open question. Unlike of a stunning computation ability to predict crystal structures with high precision, to calculate electronic and phonon structure of solid-state matter and numerically solve the Eliashberg equations within the established ME theory does provide ambiguous results for those hydrides. Dispersion of predicted T$_c$ values for particular compound is relatively large and seems to strongly depend on the used value of pseudopotential µ$^*$. It can be seen in comparison of experimental value of T$_c$ and predicted values for YH$_6$.

Results of the present study based on non-adiabatic theory of electron- vibration interaction beyond BOA offer an alternative view on theoretical aspects of superconductivity. Our study confirms that $YH_6$ and $ScH_6$ are superconductors but with an anti-adiabatic character of superconducting ground state and a multiple-gap structure in one-particle spectrum. Transition into superconducting state is driven by strong electron-phonon (EP) coupling with phonon modes of H atom vibrations. Calculated critical temperature $T_c$ in $YH_6$ is ≈ 231 K which is ≈7 K higher than the experimental value. For $ScH_6$ calculated critical temperature is $T_c$≈ 196 K, which is higher by 27 K than a former theoretical prediction ($T_c$=169 K). Unexpected results concern $CaH_6$ and $MgH_6$ in *Im-3m* structure at corresponding GPa pressures. BS calculations indicate that EP interactions in $CaH_6$ and $MgH_6$, in particular coupling to H stretching vibrations, do not induce transition from adiabatic state into superconducting anti-adiabatic ground state. These hydrides remain stable in adiabatic metal-like state characteristic for frozen equilibrium structure within the framework of Born-Oppenheimer approximation. It is in sharp contradiction to former theoretical predictions of ME theory which predicts that those two hydrides are high-$T_c$ superconductors, whose values are above 200K for $CaH_6$ and in the range of 300 K –400K for $MgH_6$.

In this respect, our results are rather indicative and cannot be understand as reference exact figures since we have employed a fairly simple semi-empirical Hamiltonian using a valence minimal atomic basis. Till this paper was submitted, high pressure syntheses of $CaH_6$, $MgH_6$ and $ScH_6$ in *Im-3m* structures have not been reported in regular journals. Nonetheless, it is worth to note that mere desirably large strengths of EP interactions and metal-like characters of the equilibrium BS are insufficient to imply superconducting properties of a solid-state matter. Topology of electronic BS and its stability with respect to optical vibration modes seems to be crucial. High pressure syntheses of $MgH_6$ or $CaH_6$ with *Im-3m* structures and experimental investigation of respective physical properties may elucidate answers related to the presented aspects.

**Notes added in reviewing process of this paper:** The authors are thankful to all the three reviewers for constructive remarks. One point should be discussed directly in this section, however. One of the reviewers directed our attention toward two manuscripts deposited recently on arXive[50,51]. These manuscripts present experimental preparation of a Ca–H system at GPa pressure level. The authors report superconductivity observed in a Ca–H system synthesized from elemental Ca and ammonia borane ($BH_3NH_3$) at 160 GPa – 180 GPa. Resistivity measurement recorded a superconductivity onset around ≈ 210 K. Structural analyses by synchrotron XRD indicates[50] that the synthesized product is a mixture of several phases and among 8 diffraction peaks, only one is consistent with a reflection of *Im-3m* $CaH_6$. Based on this fact and on the former theoretical prediction[20], the authors consider $CaH_6$ as a highly probable structure responsible for superconductivity in the mixture of synthesized Ca-hydrides. We are not in a position to interpret the reported XRD results but at the same time the presented diffraction pattern does not exclude a presence of other predicted superconducting phase[52] of Ca-H, e.g. CaH$_9$, or even a ternary phase[53] CaBH$_6$, or still some unknown superconducting phase.

Irrespective of the mentioned experimental results[50,51], we should stress some aspects related to transition into anti-adiabatic superconducting state as calculated for MH$_6$ hydrides in this paper. This concept is based on non-adiabatic theory of electron-vibration interaction beyond the BOA (see Supplementary material S1, S2 and references therein), i.e. in situation when energy scales of nuclear motion (e.g. vibrations) becomes comparable or even dominant comparing to energy scale of electronic motion in some region of the configuration space. The theory is not formulated for a specific form of an effective Hamiltonian but it was derived *ab initio* for a general non-relativistic Hamiltonian of interacting electrons and nuclei which constitute molecular or solid state systems. An electronic structure has to be calculated prior to apply the theory in investigation of a selected system. Principles of non-adiabatic theory of electron-vibration interaction are independent from the employed numerical method of the electronic structure calculation (BS in solids). Nevertheless, in some cases, the choice of a particular approach of electronic structure calculation may significantly impact the results of BS calculation and hence also the stability of BS in association to vibration modes and a possibility of anti-adiabatic state transition.

In our study, we have calculated electronic BS by using the Cyclic Cluster approach in Hartree-Fock SCF (CC HF-SCF) theory with a semiempirical INDO Hamiltonian (for more details see S2 and references therein) using minimal valence Slater-type orbitals (STO) basis sets and effective nuclear (core) charges. More explicitly, we employed 1s for $^1$H, single s and p sets for $^{12}$Mg and $^{20}$Ca, corresponding to 3s3p and 4s4p valence orbitals, respectively, whereas single s,p, and d sets were employed for $^{21}$Sc and $^{39}$Y, corresponding to 4s4p3d and 5s5p4d valence orbitals of the tow, respectively. (Let us remind that the atomic configurations are $^{12}$Mg: [Ne]$^{10}$3s$^2$; $^{20}$Ca: [Ar]$^{18}$4s$^2$; $^{21}$Sc: [Ar]$^{18}$ 4s$^2$3d$^1$; $^{39}$Y: [Kr]$^{36}$ 5s$^2$4d$^1$). In different types of superconductors (at ambient conditions) along with their non-superconducting analogues, e.g. MgB$_2$ vs. AlB$_2$, YB$_6$ vs. CaB$_6$, Nb$_3$Sn vs. Nb$_3$Sb, YBa$_2$Cu$_3$O$_7$ vs. YBa$_2$Cu$_3$O$_6$, (see references in S1) and also for compressed sulphide {(H$_2$S)$_2$,H$_2$} on GPa level[38], such basis sets appeared to be satisfactory. Inspections of BS of YH$_6$ and ScH$_6$ in Fig. 2 reveals that 5 bands which are degenerate (3+2) at Γ point in the close vicinity below FL (Fig. 2a,b – BS) are missing in BS of CaH$_6$ and MgH$_6$ (Fig. 2c,d). The pertinent 5 bands correspond to crystal orbitals combining 1s of H with five 4d functions of Y and/or 3d of Sc. Mg and Ca are "s-block" elements with negative electron affinity (EA) for Mg and a very small positive EA for Ca. In hydrides, these elements act as electron donors. At the same time those two elements belong to Group 2 of third and fourth period, while the d-orbitals with the same principal quantum numbers start to be involved in the atomic electron configurations for Sc (3d) in fourth and Y (4d) in the fifth period. Sc and Y are elements with still small EA (though relatively much higher than for Ca) and in hydrides those elements are electron donors, as well. Therefore, a legitimate question arises how an extension of the basis set would change the calculated critical parts of the BS and, in particular, to what extent involving of *d* AOs into basis sets of Mg and Ca would be crucial.

In this respect, DFT methods with using very large basis sets of plane waves (PW) are superior in studies structural aspects related to search for minimum total energy of solid state matter. However, these approaches are not free from ambiguities, too. It is well known that eigenenergies $\{\varepsilon_i\}$ of Kohn-Sham equations have no real physical meaning, i.e., in contrast to Hartree-Fock eigenvalue solution, $\{\varepsilon_i\}$ are not orbital energies. Beside contributions of $\varepsilon_i$ and Hartree energy, the total energy of a system within DFT theory is mainly determined by contributions related to exchange-correlation potential ($v_{xc}$). Well known is fact that for small gap semiconductors, calculated value of gap (by means of energy difference of highest occupied orbital energy $\varepsilon_i$ and lowest unoccupied orbital energy $\varepsilon_a$ at FL) strongly depends on the choice of $v_{xc}$ potential. Nonetheless, set of Kohn-Sham eigenenergies $\{\varepsilon_i\}$ is routinely employed in electronic BS calculation of superconductors with PBE exchange-correlation potential. While this potential is widely used mainly for structure optimization (total energy criterion), it is less effective for band gap calculation, i.e. for calculation of $\{\varepsilon_i\}$ close to Fermi level. Predictions of superconductivity in $CaH_6$[20] and in $MgH_6$[22] are based on DFT calculations with PBE exchange-correlation potential in PW basis set. Inspection of BS for $CaH_6$ in the original work[20] (see Figure S17 in Ref[20]) show bands that are degenerated at Γ point slightly below FL which appear similar to "d-bands" degenerated in the vicinity of FL in BS of $YH_6$ and $ScH_6$ in this work – Figure 2a,b. Projection (APW) on radial part of atomic orbitals indicates that these bands in Figure S17-Ref[20] are characteristic by combination of H-1s and Ca-3d (see Figure S16 in Ref[20] with PDOS decomposition close to FL). Prompted by this fact, we have performed a quick check when we have confirmed the d- character of these bands also directly by DFT test calculations with STO basis set (ADF package). Single Zeta (SZ) basis set can be compared to our valence minimal basis set. With this basis, the DFT/PBE calculation resulted in the same topology of the BS as in our calculations for both Ca and Mg (Fig. 2c,d), i. e. the bands of d-character that are degenerated at Γ point slightly below the FL are missing. On the other hand, more expanded basis sets, e.g. double and triple zeta with polarization functions (DZP, TZP) that include 3d orbitals for Ca, provide BS topologies of the same features as in Figure S17 of Ref.[20], i. e. the degenerated "d-bands" just below the FL at Γ point, and, in association with stretching H-vibration these bands start to fluctuate.

To express it explicitly, for $CaH_6$, DFT/PBE approaches with "large" PW basis set or d-containing STO basis set provide electronic BS of a same topological character in Γ point at FL as BS calculated for hydrides of "d-group" elements $YH_6$ and $ScH_6$. This type of band structure is unstable in coupling to H-stretching vibration and is responsible for high-$T_c$ superconductivity in superhydrides of "d-group" metal elements $YH_6$ and $ScH_6$. In a formal way, based on the topological character of BS, superconducting should also be $CaH_6$. We have not tested $MgH_6$ in detail but, in this case, the degenerated d-bands appear also close to FL in Γ point and probably BS is also unstable with respect to H-stretching vibration.

In relation with a role of d-electrons in the possible superconductivity of $CaH_6$ and/or $MgH_6$, it is worth noting that the energy cost[54] for a promotion of 4s electron to 3d in Ca is 2.71 eV, i.e. for a transition $^1S$ $(4s^2)$ state to $^1D$ $(3d^14s^1)$, whereas similar transition energy in Mg ($^1S$ $3s^2$ → $^1D$ 3s3d) is substantially higher (5.75 eV). Whether the GPa pressure levels enable/justify such electronic transitions in the involved elements remains to be further investigated.

Legitimate question arises whether the results presented in this work for $CaH_6$ and $MgH_6$ should mean the failure of the anti-adiabatic theory. The answer is no. These results do not concern the theory of superconductivity at all. The discrepancies are related to technical problems in calculation of electronic BS as aforementioned. Nevertheless, these are serious problems that deserve a separate investigation. We believe that question of superconductivity in $CaH_6$ and $MgH_6$ will be resolved sooner or later by successful synthesis of pure Im-3m phase of these compounds on GPa level.

**Supplementary material**

Basic notes on non-adiabatic theory of electron –vibration interactions are presented in part S1. In part S2 are information on CC HF-SCF method and relevant equations of anti-adiabatic theory.

**Acknowledgement**

P.B. thanks his family for long-term support and understanding. J.N. is grateful to the Grant Agency of the Ministry of Education of the Slovak Republic and Slovak Academy of Sciences (VEGA project No. 1/0712/18) as well as to the Slovak Research and Development Agency (APVV-17-0324) for support of this work.

**Data availability**

Data that support the findings of this study are available within the article and its Supplementary material.

# References

\* Author to whom correspondence should be addressed, banacky@fns.uniba.sk; or  47palob@gmail.com

1. a/ A.R. Oganov,  C.W. Glass,  *J. Chem. Phys.* 124,244704 (2006)
   b/ A.R. Oganov, A.O. Lyakhov and M. Valle,  Acc. Chem. Res. 44, 227 (2011)
   c/ A.O. Lyakhov,  A.R. Oganov,  H.T.Stokes,  Q. Zhu,  Comput. Phys. Commun.184, 1172  (2013)
2. C.J. Pickard, R.J. Needs,  *Phys. Rev. Lett.* 97, 045504   (2006)
3. a/ Y. Wang, J. Lv, L. Zhu, Y. Ma,  *Phys. Rev. B. 82*, 094116  (2010) ;
   b/ Y. Wang, J. Lv, L. Zhu  and Y. Ma, Comput. Phys. Commun. **183**, 2063 (2012)
4. P. Hohenberg , W. Kohn, *Phys. Rev.* 136, B864 (1964)
5. W. Kohn,  L.J. Sham, *Phys. Rev.* 140, A1133 (1965)
6. a/ G. Kresse and  J. Furthmuller,  Phys. Rev. B 54, 11169 (1996)
   b/ G. Kresse and D. Joubert,  Phys. Rev. B 59, 1758  (1999)
   c/ P.E. Blöchl, Phys. Rev. B **50**, 17953 (1994)
   d/ G. Kresse and J. Furthmuller, *Comput. Mater. Sci.* 6(1), 15 (1996)
7. Y. Wang and Y. Ma, J. Chem. Phys., 2014, 140, 040901
8. P. Giannozzi, S. Baroni, N. Bonini, M. Calandra, R. Car, C. Cavazzoni, D. Ceresoli, G. L Chiarotti, M. Cococcioni, I. Dabo, A. Dal Corso, S. De Gironcoli, S. Fabris, G. Fratesi, R. Gebauer, U. Gerstmann, C. Gougoussis, A. Kokalj, M. Lazzeri, L. Martin-Samos, N. Marzari, F. Mauri, R. Mazzarello, S. Paolini, A. Pasquarello, L. Paulatto, C. Sbraccia, S. Scandolo, G. Sclauzero, A. P. Seitsonen, A. Smogunov, P. Umari, R. M. Wentzcovitch, QUANTUM ESPRESSO: a modular and open-source software project for quantum simulations of materials, J. Phys.: Condens. Matter 21,395502  (2009)
9. W. L. McMillan,  Phys. Rev. 167, 331 (1968)
10. P. B. Allen and R. C. Dynes, Phys. Rev. B 12, 905 (1975)
11. R. C. Dynes, Solid State Commun. 10, 615 (1972).
12. A. B. Migdal, Zh. Eksp. Teor. Fiz. 34,1438 (1958);  Sov. Phys. JETP 7, 996 (1958)
13. G. M. Eliashberg, Zh. Eksp. Teor. Fiz. 38, 996 (1960);  39,1437 (1960);  Sov. Phys. JETP 11 696 (1960); 12,1000 (1960)
14. J. Bardeen, L.N. Cooper, and J.R. Schrieffer, Phys. Rev. 108, 1175 (1957)
15. F. Marsiglio, J.P. Carbotte,  Electron-phonon superconductivity, in Superconductivity,  Springer, Berlin, Heidelberg, pgs.73-162  (2008)
16. C. J. Pickard,  I. Errea and M. I. Eremets,  Ann. Rev. Cond. Matter Physics  11, 57 (2020)
17. N.W. Ashcroft, *Phys. Rev. Lett.* 21, 1748 (1968);  Phys. Rev. Lett. **92**, 187002 (2004)
18. A.P. Drozdov, M.I. Eremets, I.A. Troyan, V. Ksenofontov,  S. I. Shylin, Nature  525, 73 (2015)
19. D. Duan, Y. Liu, F. Tian, D. Li, X. Huang, Z. Zhao, H. Yu, B. Liu, W. Tian and T. Cui,  Sci Rep 4, 6968 (2014)
20. H. Wang, J. S. Tse, K. Tanaka, T. Iitaka  and Y. Ma,  Proc. Natl. Acad. Sci. USA 109, 6463 (2012)
21. K. Tanaka, J. S. Tse  and H. Liu,  Phys. Rev. B  96, 100502(R)  (2017)
22. X. Feng, J. Zhang, G. Gao, H. Liu  and H. Wang,  RSC Adv. **5**, 59292 (2015)
23. R. Szcześniak and A. P. Durajski, Front. Phys. 11(6), 117406 (2016)
24. F. Peng , Y. Sun , C.J Pickard , R.J Needs , Q. Wu , Y. Ma, Phys Rev Lett  119, 107001 (2017).

25. H. Liu, I. I. Naumov, Z. M. Geballe, M. Somayazulu, J. S.Tse and R. J. Hemley Phys. Rev. B 98, 100102(R) (2018)
26. H. Liu, I.I. Naumov, R. Hoffmann, N.W. Ashcroft and R.J. Hemley, Proc. Natl Acad. Sci. USA 114, 6990–6995 (2017).
27. A.P. Drozdov, P.P. Kong, V.S. Minkov, S.P. Besedin, M.A. Kuzovnikov, S. Mozaffari, L. Balicas, F.F. Balakirev, D.E. Graf, V.B. Prakapenka, E. Greenberg, D.A. Knyazev, M. Tkacz and M.I. Eremets, Nature 569, 528 (2019)

28. E. Snider, N. Dasenbrock-Gammon, R. McBride, M. Debessai, H. Vindana, K. Vencatasamy, K. V. Lawler, A. Salamat , R. P. Dias, Nature 586**,** 373 (2020), https://doi.org/10.1038/s41586-020-2801-z

29. a/ I.A. Troyan, D.V. Semenok, A.G. Kvashnin, A.G. Ivanova, V.B. Prakapenka, E. Greenberg, A.G. Gavriliuk, I.S. Lyubutin, V.V. Struzhkin and A.R. Oganov, *arXiv:1908.01534 v1* (2019);
b/ A.G. Kvashnin, A.V. Sadakov, O.A. Sobolevskiy, V.M. Pudalov, A.G. Ivanova, V.B. Prakapenka, E.Greenberg, A.G. Gavriliuk, V.V. Struzhkin, A. Bergara, I. Errea, R. Bianco, M. Calandra, F. Mauri, L. Monacelli, R. Akashi and A.R. Oganov, *arXiv:1908.01534v2* (2019)

30. P.P. Kong, V.S. Minkov, M.A. Kuzovnikov, S. P. Besedin, A.P. Drozdov, S. Mozaffari, L.Balicas, F.F. Balakirev, V.B. Prakapenka, E. Greenberg, D.A. Knyazev and M. I. Eremets, *arXiv:1909.10482* (2019)

31. C. Heil, S. di Cataldo, G.B. Bachelet and I. Boeri, Phys. Rev. B 99, 220502 (2019)

32. Y. Li, J. Hao, H. Liu, J.S. Tse, Y. Wang and Y. Ma , Scientific Reports **5**, 09948 (2015)

33. H. Liu, I.I. Naumov, R. Hoffmann, N.W. Ashcroft and R.J. Hemley, PNAS 114, 6990 (2017)

34. K. Tanaka, J.S. Tse and H. Liu, Phys.Rev.B ; 96,100502(R) (2017)

35. K.S. Grishakov, N.N. Degtyarenko and E.A. Mazur, J. Exp. Theor. Phys. 128, 105 (2019).

36. K. Abe, Phys.Rev.B 96,144108 (2017)

37. X. Ye, N. Zarifi, E. Zurek, R. Hoffmann, and N. W. Ashcroft, J. Phys. Chem. C 122, 6298 (2018)

38. P. Baňacký, Results in Physics 6,1 (2016)

39. a/ L. Pietronero, Europhys. Lett. 17, 365 (1992)
b/ L. Pietronero and S. Strassler, Europhys. Lett. 18, 627 (1992)
c/ L.Pietronero, S.Strassler and C.Grimaldi, Phys.Rev.B 52, 10516 (1995)

40. a/ G.Grimaldi, L. Pietronero and S. Strassler, Phys. Rev. B52, 10530 (1995)
b/ G.Grimaldi, L. Pietronero and S. Strassler, Phys. Rev. Lett. 75, 1158 (1995)

41. a/ G.Grimaldi, E.Cappelluti and L. Pietronero, Europhys. Lett. 42 , 667 (1998)
    b/ G.Grimaldi, L. Pietronero and M.Scottani, Europhys. J. B 10, 247 (1999)
    c/ E.Cappelluti, G.Grimaldi and L. Pietronero, Phys. Rev. B64, 125104 (2001)
    d/ M.Botti, E.Cappelluti, G.Grimaldi and L. Pietronero, Phys. Rev. B66, 054532 (2002)
    e/ E.Cappelluti, S.Ciuchi, G.Grimaldi and L. Pietronero, Phys. Rev. B68, 174509 (2003)

42. a/ A.S. Alexandrov and J. Runninger, Phys. Rev. B 23, 1796 (1981)
    b/ A.S. Alexandrov and J. Runninger, Phys. Rev. B 24, 1164 (1981)
    c/ A.S. Alexandrov, Zh.Fiz.Chim 57, 273 (1983) (Rus. J.Phys.Chem. 57, 167 (1983)

43. A.S. Alexandrov, Physica C 363, 231 (2001)

44. a/ A.S. Alexandrov, Theory of superconductivity: From weak to strong coupling (IoP Publishing, Bristol and Philadelphia, 2003)
    b/ A.S. Alexandrov, arXiv:cond-mat/0508769v4 (2005)

45. A.S. Alexandrov and P.P. Edwards, Physica C 331, 97 (2000)

46. A.S. Alexandrov, Phys. Rev. B 23, 2838 (1992)

47. J.P. Hague, P.E. Kornilovich, J.H. Samson and A.S. Alexandrov, J.Phys - Cond.Matt. 19, 255214 (2007)

48. P. Baňacký, Int. J. Quant. Chem.,101, 131(2005)

49. L.Boeri, E. Cappelluti and L.Pietronero, Phys.Rev.B, 71, 012501 (2005)

50. L. Ma, K. Wang, Y. Xie, X. Yang, Y. Wang, M. Zhou, H. Liu, G. Liu, H. Wang and Y. Ma, arXiv:2103.16282, (2021)

51. Z.W. Li, X. He, C.L. Zhang, S.J. Zhang, S. M. Feng, X.C. Wang, R.C. Yu, C.Q. Jin, arXiv: 2103.16282v1 (2021)

52. Z. Shao, D. Duan, Y. Ma, H. Yu, H. Song, H. Xie, D. Li,F. Tian, B. Liu, and T. Cui, Inorganic chemistry58, 2558 (2019).

53. S. di Cataldo, W. von der Linden, L. Boeri, arXiv:2006.00960v1 (2020)

54. **https://physics.nist.gov/PhysRefData/Handbook/Tables/calciumtable5.htm**

# Supplementary material

**Aspects of strong electron-phonon coupling in superconductivity of compressed metal hydrides MH$_6$ (M=Ca, Mg, Y, Sc) with *Im-3m* structure**

Pavol Baňacký and Jozef Noga

Comenius University, Faculty of Natural Sciences, Chemical Physics group, Department of Inorganic Chemistry, Mlynska dolina CH2, 84215 Bratislava, Slovakia

## S1. Notes to the non-adiabatic theory of electron –vibration interactions

Non-adiabatic theory of electron–vibration interactions[1] is an extension of standard adiabatic BOA where electronic motion is parametrically dependent on fixed nuclear coordinates. In non-adiabatic extension beyond the BOA, electronic motion is introduced to be explicitly dependent on the operators of instantaneous nuclear coordinates and also on the operators of instantaneous nuclear velocities–momenta. For a non-adiabatic regime, $\omega/E_F \approx 1$, with comparable values of nuclear and electronic velocities, or for anti-adiabatic regime $\omega/E_F > 1$, the electronic subsystem influences not only nuclear potential energy, but also kinetic energy of the nuclear motion of involved atoms is changed for some degrees of freedom. Under these circumstances, direct–independent solution (diagonalization) of the electronic Schrodinger equation is impossible. We have solved this problem by sequence of canonical transformations of general form of "molecular" Hamiltonian. Formulation of non-adiabatic theory of electron-vibration interactions beyond BOA has been published[1] in series of papers in theoretical chemistry oriented Int. J. Quantum Chemistry in 1992. The theory was verified by high-precision quantum chemical *ab-initio* calculations of non-adiabatic energy correction to standard adiabatic total electronic energy calculated for systems in fixed nuclear geometry. Non-adiabatic corrections account for nuclear motion and can be decomposed on contributions of particular normal modes, i.e. vibration and rotation/translation modes. With spectroscopic precision, small molecular systems have been studied[2]. More specific arrangement of the theory toward superconductivity is available in the form of open access papers[3-5] in physical journals. On semiempirical level, the theory was applied to study superconductivity[6-13] in anti-adiabatic state for different groups of superconductors or to study some spectral properties of superconductors, e.g. ARPES[6,7] or pump-probe spectroscopy of relaxation processes in superconductors[13]. It has been shown that, no matter if EPC is strong or weak, adiabatic → anti-adiabatic state transition is the effect common for superconductors of different groups, e.g. classical metal alloys - Nb$_3$Sn, Nb$_3$Ge,…, covalent MgB$_2$, YB$_6$, high-Tc cuprates - e.g. YBa$_2$Cu$_3$O$_7$, and also for predicted superconductors in single and multi-wall B$\alpha$8 nanotubes with and without metal doping. Effect of adiabatic → anti-adiabatic state transition is absent in corresponding non-superconducting analogues, e.g. Nb$_3$Sb, CaB$_6$, AlB$_2$, YBa$_2$Cu$_3$O$_6$. In contrast to BCS with Cooper pairs condensation in k -space, according to anti-adiabatic theory due to the effect of nuclear dynamics, the system is stabilized as an anti-adiabatic state of broken symmetry with a gap(s) in its one-particle spectrum. Distorted nuclear structure related to nuclear displacement in the phonon mode *r* which induces transition of system in anti-adiabatic state, has fluxional character. Geometric degeneracy (i.e. fluxional structure) of the anti-adiabatic ground state enables formation of mobile bipolarons in real space. Bipolarons, in form of inter-site polarized charge density distribution, can move over lattice in external electric potential as super-carriers without dissipation. Thermodynamic properties[3-5] of system in stabilized anti-adiabatic state corresponds to thermodynamics of superconductors. Necessary equations for study superconductivity in $MH_6$ hydrides are presented in S2.

## S2. Methods

For theoretical study of potential superconductivity in a solid state compound, crucial is investigation of electronic band structure (BS). Important is topology of BS in frozen-fixed equilibrium geometry and stability of BS topology in EP interactions. Superconductors in equilibrium structure are characteristic by metal-like BS, with very small (or small) adiabatic ratio, $\omega/E_F \ll 1$ (or $\omega/E_F \lesssim 1$). If EPC induces BS instability to such extend that the adiabatic BOA is broken, i.e. original adiabatic $\omega/E_F$ is switched to anti-adiabatic regime, $\omega/E_F > 1$, then application of anti-adiabatic theory beyond the BOA can disclose potential superconductivity of studied system. To verify if a compound is prone to adiabatic →anti-adiabatic state transition, frozen phonon mode method is used. Here, the BS is calculated for a set of fixed atom displacements $\delta$ out–off equilibrium in particular phonon mode. According to basic output of the BOA, which enables separate description and solution of electronic and nuclear subsystems of molecular or solid state system, the wave function can be factorized in the form (S2.1) only if inequality $\omega/E_F \ll 1$ (or at least $\omega/E_F \lesssim 1$) holds, i.e. system is in adiabatic state;

$$\Psi(r, Q) = \Phi(r, [Q])X([Q]) \tag{S2.1}$$

Factorized form (S2.1) enables to study electronic subsystem $\Phi(r,[Q])$ for a fixed set of nuclear coordinates $[Q]$. Nuclear coordinate configuration $[Q]$ is only parameter in solution of electronic Schrödinger equation. Electronic energies $E_i([Q]_i)$ for different fixed sets $[Q]_i$ (including $[Q]_{eq}$) constitute potential energy of nuclear motion. For system to stay in adiabatic state, i.e. for validity of BOA, factorized form (S2.1) must hold not only for $[Q]_{eq}$ but also for relevant configuration space $\{[Q]\}$ of nuclear motion. Relevant configuration space of nuclear motion has to cover at least displacements of atoms out-off equilibrium for at least ground vibration state (zero-point energy) of high-frequency phonons in a system. Such critical displacements can be calculated by numerical solution of nuclear Schrödinger equation for particular vibration mode. In case that for given mode contributions of anharmonicity are not crucial, an approximate treatment within linear harmonic oscillator (LHO) approximation is justified. The root-mean-square displacement ($q_{r.m.s}$) can be derived on the LHO level in the analytical form (S2.2);

$$q_{r.m.s.} = \frac{1}{2}\left(\frac{\langle H \rangle}{m\omega^2}\right)^{1/2} \tag{S2.2}$$

The mean value of the oscillator is, $\langle H \rangle = -\dfrac{\partial}{\partial \beta}\left(\ln Tr(e^{-\beta H})\right) = \dfrac{1}{2}\hbar\omega \coth\left(\dfrac{1}{2}\beta\hbar\omega\right)$

In anti-adiabatic state, electronic subsystem cannot be studied as separate subsystem for a set of fixed nuclear coordinates [Q]. In this case, electronic subsystem has to be studied as fully dependent on instantaneous nuclear coordinates as well dependent also on instantaneous nuclear velocities. Solution of this problem is covered by theory of non-adiabatic electron-vibration interactions[1]. In present study we apply derived equations rearranged into form of k-space representation[3-5] suitable to study periodic solids. It enables to calculate corrections to total electronic energy ($\Delta E^0_{(na)}$), corrections to adiabatic orbital (band) energies near FL and corrected DOS (PDOS) near FL induced by vibration motion of nuclei in stretching mode $\omega_r$ which in EPC drives crossing of adiabatic system into anti-adiabatic state.

Corrections to adiabatic ground state total electronic energy ($\Delta E^0_{(na)}$) in *k*-space representation is,

$$\Delta E^0_{(na)} = 2 \sum_{\varphi_{Rk}} \sum_{\varphi_{Sk'}} \int_0^{\varepsilon_{k',\max}} n^0_{\varepsilon_{k'}} \left(1 - f_{\varepsilon^0_{k'}}\right) d\varepsilon^0_{k'} \int_{\varepsilon_{k,\min}}^{\varepsilon_{k,\max}} f_{\varepsilon^0_k} \left|u^r_{k-k'}\right|^2 n^0_{\varepsilon_k} \frac{\hbar \omega_r}{\left(\varepsilon^0_k - \varepsilon^0_{k'}\right)^2 - \left(\hbar \omega_r\right)^2} d\varepsilon^0_k, \quad \varphi_{Rk} \neq \varphi_{Sk'} \qquad (S2.3)$$

Terms $\varepsilon^0_k < \varepsilon_F$; $\varepsilon^0_{k'} > \varepsilon_F$ are adiabatic values of electronic band energies in *k* point below FL and in *k'* point above FL for system with fixed nuclear position in equilibrium. Fermi-Dirac populations $f_{\varepsilon^0_k}$, $f_{\varepsilon^0_{k'}}$ make correction (S2.3) temperature-dependent. Term $u^r_{k-k'}$ stands for matrix element of EP coupling and $n^0_{\varepsilon_{k'}}$, $n^0_{\varepsilon_k}$ are PDOS of interacting bands at $\varepsilon^0_{k'}$ and $\varepsilon^0_k$. For adiabatic systems, i.e. basically for all molecular systems or solid-state metals, insulators and semiconductors, this correction is small and positive, and represents negligible diagonal Born-Oppenheimer correction (DBOC) to total electronic energy. Only for systems in anti-adiabatic state when $\omega/E_F > 1$, **correction $\Delta E^0_{(na)}$ can be negative** and its absolute value depends on the magnitudes of $u^r_{k-k'}$ and $n^0_{\varepsilon_{k'}}$, $n^0_{\varepsilon_k}$ for distorted structure with particular vibration displacement $\delta$ when analytical critical point (ACP) of fluctuating band approaches FL. In these circumstances the system not only undergoes transition to an anti-adiabatic state but PDOS of fluctuating band is considerably increased and resembles van Hove singularity close to FL. Whether system remains stabilized in anti-adiabatic ground state with distorted structure depends on absolute value of $\Delta E^0_{(na)}$. On adiabatic level the ground state is reached for equilibrium geometry with frozen nuclear position – high symmetry structure. Any nuclear displacements out-off equilibrium decreases symmetry and increases (destabilizes) total electronic energy of system by $\Delta E_\delta > 0$ (contribution of nuclear potential energy at displacement $\delta$). However, if for critical $\delta_{cr}$, $abs(\Delta E^0_{(na)}) > \Delta E_{\delta cr}$, then electronic system is stabilized in anti-adiabatic ground state with distorted geometry. Anti-adiabatic stabilization energy, $\Delta E_{(aa)} < 0$, is equal to $\Delta E_{(aa)} = \Delta E^0_{(na)} + \Delta E_{\delta cr}$. We have shown elsewhere[48] that in this way stabilized anti-adiabatic ground state is geometrically degenerated with fluxional nuclear structure. For system stabilized in anti-adiabatic ground state, involved nuclei displaced by $\delta_{cr}$ (i.e. critical value for shift of ACP of fluctuating band across FL) can revolve in concert on perimeters of circles with radius $\delta_{cr}$ and with centers in adiabatic equilibrium positions without dissipation, while total electronic energy remains unchanged. Geometric degeneracy of anti-adiabatic ground state and stabilization energy $\Delta E_{(aa)}$ enable BEC in a form of mobile bipolaron formation in real space (i.e. form of mobile inter-site polarized charge density distribution) due to translational symmetry of lattice. In this sense, it is real space counterpart of Cooper pair condensation energy in k-space of BCS theory.

In anti-adiabatic state, $k$-dependent gap $\Delta_k(T)$ in quasi-continuum of adiabatic metal-like one-electron spectrum is opened. Gap opening is related to shift $\Delta\varepsilon_{Pk}$ of equilibrium adiabatic band energies $\varepsilon^0_{Pk}$, $\varepsilon_{Pk} = \varepsilon^0_{Pk} + \Delta\varepsilon_{Pk}$, and to the $k$-dependent change of PDOS in particular band(s) close FL. Shift of orbital energies in band $\varphi_P(k)$ has the form[3-5],

$$\Delta\varepsilon(Pk') = \sum_{Rk'_1 > k_F} |u^{k'-k'_1}|^2 (1 - f_{\varepsilon^0 k'_1}) \frac{\hbar\omega_{k'-k'_1}}{(\varepsilon^0_{k'} - \varepsilon^0_{k'_1})^2 - (\hbar\omega_{k'-k'_1})^2} - \sum_{Sk < k_F} |u^{k-k'}|^2 f_{\varepsilon^0 k} \frac{\hbar\omega_{k-k'}}{(\varepsilon^0_{k'} - \varepsilon^0_{k})^2 - (\hbar\omega_{k-k'})^2} \qquad (S2.4)$$

for $k' > k_F$, $u^{k'-k'_1} = u^q_{PR}$ and

$$\Delta\varepsilon(Pk) = \sum_{Rk'_1 > k_F} |u^{k-k'_1}|^2 (1 - f_{\varepsilon^0 k'_1}) \frac{\hbar\omega_{k-k'_1}}{(\varepsilon^0_k - \varepsilon^0_{k'_1})^2 - (\hbar\omega_{k-k'_1})^2} - \sum_{Sk_1 < k_F} |u^{k-k_1}|^2 f_{\varepsilon^0 k} \frac{\hbar\omega_{k-k_1}}{(\varepsilon^0_k - \varepsilon^0_{k_1})^2 - (\hbar\omega_{k-k_1})^2} \qquad (S2.5)$$

for $k \leq k_F$

Replacement of discrete summation by integration, $\sum_{k}... \to \int n^0_{\varepsilon_k}$, introduces PDOS $n^0_{\varepsilon_k}$, $n^0_{\varepsilon_{k'}}$ into Eqs (S2.4,5) and, which is of crucial importance in relation to fluctuating band - see Fig 3b, 4b in the main text. Shift $\Delta\varepsilon_k$ of original adiabatic band energies outside from FL ($\varepsilon^0_k$ are shifted more below FL and $\varepsilon^0_{k'}$ are shifted more above FL) corrects also original adiabatic PDOS $n^0(\varepsilon_k)$. For corrected DOS, following relation has been derived[6];

$$n(\varepsilon_k) = |1 + (\partial(\Delta\varepsilon_k)/\partial\varepsilon^0_k)|^{-1} n^0_{\varepsilon_k} \qquad (S2.6)$$

Term $n^0_{\varepsilon_k}$ stands for uncorrected PDOS of the original adiabatic band-states in equilibrium,

$$n^0_{\varepsilon_k} = |(\partial\varepsilon^0_k/\partial k)|^{-1} \qquad (S2.7)$$

Close to the $k$-point where the original band that interacts with fluctuating band intersects FL, the occupied states near FL are shifted downward below FL and unoccupied states are shifted upward - above FL. The gap is identified as the energy distance between created peaks in corrected DOS above FL (half-gap) and below FL. The formation of peaks is related to the spectral weight transfer that is observed in superconductors by ARPES or tunneling spectroscopy in cooling below $T_c$. Opened gap in one-particle spectrum was derived[1,3-5] in the form (S2.8),

$$\Delta(T) = \Delta(0) tgh\left(\Delta(T)/4k_B T\right) \tag{S2.8}$$

Corrections to orbital energies (S2.4, 5) and to the ground state energy (S2.3) are temperature dependent and decrease with increasing T. At a critical value $T_c$, the gap in one-particle spectrum (S2.8) disappears - $\Delta(T_c) = 0$ and continuum of band states is established at FL. In these circumstances hold $\left|\Delta E^0_{(na)}(T_c)\right| \leq \Delta E_{d,cr}$ and the system undergoes transition from the anti-adiabatic into adiabatic state, which is stable for equilibrium high-symmetry structure above $T_c$. With respect to $\Delta(0)$, simple approximate relation follows from equation (S2.8),

$$T_c = \Delta(0)/4k_B \tag{S2.9}$$

Correction to the ground state total electronic energy (S2.3) is in adiabatic state reduced to small positive value,

$$\Delta E^0_{(a)}(T > T_c) \approx \sum_{qk} |u^q|^2 \frac{\hbar \omega_r}{\left(\varepsilon^0_k - \varepsilon^0_{k-q}\right)^2} \geq 0 \tag{S2.10}$$

Expression (S2.10) is basically an energy contribution of standard adiabatic polaron to the total energy of system in adiabatic state[1,3].

Electronic band structure (BS) of studied hydrides in fixed nuclear geometries has been calculated by a computer code[14] Solid2000. This code is not based on Kohn-Sham DFT method but on the Hartree-Fock SCF (HF-SCF) method for infinite 3D-periodic cyclic cluster (CC)[14] with semiempirical quasi-relativistic INDO Hamiltonian[15] and parameterization based on results of atomic Dirac-Fock calculations[16]. This method of BS calculation has many advantages and some drawbacks as well. The HF-SCF CC method is inconvenient for strong ionic crystals but yields good results for intermediate ionic and covalent systems. The main disadvantage is overestimation of the total bandwidth, but it yields satisfactory results for properties related to electrons close to Fermi level (frontier-orbital properties) and for equilibrium geometries[17-19]. In practical calculations, the basic cluster of the dimension ($N_a$ x $N_b$ x $N_c$), is generated by corresponding translations of the unit cell in the directions of crystallographic axes, *a* ($N_a$), *b* ($N_b$), *c* ($N_c$), with imposed Born-von Karman boundary conditions. "Infinite" – 3D-periodic cyclic cluster structure is generated in calculation of matrix elements. In particular for studied $MH_6$ hydrides band structure calculations have been performed for the basic clusters 11x11x11. Scaling parameter 1.2 has been used in calculations of the one-electron off-diagonal two-center matrix elements of the Hamiltonian (i.e., β-"hopping" integrals). The basic cluster of a given size generates a grid of ($N_a$ x $N_b$ x $N_c$) points in *k*-space. For $MH_6$ it is 1331 k-points. The HF-SCF CC procedure is performed for each k-point of the grid with the INDO Hamiltonian matrix elements that obey the boundary conditions of the cyclic cluster[14]. The Pyykko-Lohr quasi-relativistic basis set of the valence electron atomic orbitals AO (s,p-AO for Mg, Ca ; s,p,d-AO for Sc, Y and 1s for H atom) has been used. The number of Slater type AO`, i.e. STO-type functions, is unambiguously determined by number of valence AOs pertaining to atoms of basic cluster. In general, the precision of the results of band structure calculation increases with increasing dimension of the basic cluster. It has been shown[14, 17-19], however, that there is an effect of saturation, a bulk limit beyond which the effect of increasing dimension on e.g. total electronic energy, orbital energies, HOMO-LUMO difference…, is negligible. In practice, the dimension of the basic cluster and parameter selection (e.g. for calculation of β integrals) is a matter of compromise between computational efficiency and convergence of calculated electronic properties and equilibrium geometry to some reference or experimental data. Note, however, that the basic efficiency and accuracy are restricted by the INDO parameterization. Important is that besides total electronic energy which cover explicitly also contributions of two-electrons term (e-e interactions), extracted can be contribution of one-electron term, $h_{1PR}(r, [Q])$, which comprise besides electron kinetic energy also electron-nuclear (ion) interaction at frozen geometry. On an approximate level, performing calculations for a set of fixed but displaced configurations (with generalized coordinate Q) in vibration mode *(i)* enable to calculate electron-vibration interaction;

$$h_{1PR}(r, Q) = h_{1PR}(r, [Q_0]_{eq}) + \sum_{i=1}^{\infty} u_{PR}^{(q)}\{Q\} \quad u_{PR}^q(Q) = \frac{\partial h_{1PR}(Q)}{\partial Q^{(i)}} Q^{(i)} = u_{PR}^q Q^{(i)} \tag{S2.11}$$

In this study, for electron vibration interactions, instead of matrix elements $u_{PR}^{(q)}$ for mode *(i)* and electronic states *{P, R}* we use a mean value,

$$g_{\delta_{cr}} = \left( \frac{\partial \langle h_1(Q) \rangle}{\partial Q^{(i)}} \right)_{Q_{\delta_{cr}}} Q^{(i)} \tag{S2.12}$$

Reason is pragmatic - semiempirical character of INDO Hamiltonian, i.e. unless all 1 and 2-electron integrals of system Hamiltonian are not calculated on *ab initio* level, $g_{cr}$ value is reasonable for present study.

**Reference**s [a]

[a] References for Supplementary Material

1. M. Svrček, P.Baňacký and A.Zajac, Int.J.Quant.Chem. 43,393 (1992); Int.J.Quant.Chem 43, 415 (1992); Int.J.Quant.Chem  43, 425 (1992); Int.J.Quant.Chem 43, 551 (1992)
2. M. Svrček, P.Baňacký, S.Biskupič, J.Noga, P.Pelikan and A.Zajac, Chem. Phys. Lett. 299, 151 (1999)
3. P.Baňacký, Adv.Cond.Matter Physics, 2010,752943 (2010), https://doi.org/10.1155/2010/752943
4. P.Baňacký,  J.Phys.Chem.Solids, 69, 2728 (2008)
5. P.Baňacký,  J.Phys.Chem.Solids, 69, 2696 (2008)

6. P. Baňacký, M. Svrček, V. Szöcs, Int. J. Quant. Chem., 58 (5), 487 (1996)
7. P. Baňacký, in Superconducting Cuprates: Properties, Preparation and Applications, pgs. 187-212, New York, Publ. Nova Science (2009), ISBN 978-1-60456-919-3
8. P. Baňacký, in Progress in Theoretical Chemistry and Physics Vol. 22, Advances in the Theory of Quantum Systems in Chemistry and Physics, pgs. 483-512, Dordrecht, Springer (2011), ISBN 978-94-007-2076-3
9. P. Baňacký, Collection Czechoslovak Chemical Communications, 73(6-7), 795 (2008)
10. P. Baňacký, J. Noga, V. Szöcs, J Phys Chem Solids, 73 (8), 1044 (2012)
11. P. Baňacký, P. Noga, V. Szöcs, J. Noga, Physica Status Solidi B-Basic Solid State Physics, 252(9), 2052 (2015)
12. P. Baňacký, J. Noga, V. Szöcs, Quantum Matter 4(4), 367 (2015)
13. P. Baňacký, V. Szöcs, Physica Status Solidi B - Basic Solid State Physics., 253(5), 942 (2016)
14. J. Noga, P. Baňacký, S. Biskupič, R. Boča, P. Pelikan, and A. Zajac, J. Comp. Chem. 20, 253 (1999)
15. J.A. Pople and D.L. Beveridge, in Approximate Molecular Orbital Theory, McGraw-Hill Inc, New York, (1970)
16. R. Boča, Int. J. Quant. Chem. 31, 941 (1988); Int. J. Quant. Chem. 34, 385 (1988)
17. A. Zajac, P. Pelikán, J. Noga, P. Baňacký, S. Biskupič and M. Svrček, J. Phys. Chem. B 104, 1708 (2000)
18. A. Zajac, P. Pelikán, J. Minar, J. Noga, M. Straka, P. Baňacký and S. Biskupič, J. Solid. State Chem. 150, 286 (2000)
19. P. Pelikán, M. Kosuth, S. Biskupič, J. Noga, M. Straka, A. Zajac and P. Baňacký, Int. J. Quant. Chem. 84, 157 (2001)